\documentclass[twocolumn,prb]{revtex4}
\usepackage{amsfonts}
\usepackage[T1]{fontenc}
\usepackage{amsmath,amsbsy,amssymb,graphicx}
\usepackage{times}

\let\mathbf=\boldsymbol

\begin{document}

\title{Non-Hermitian boundary and interface states\\
in nonreciprocal higher-order topological metals and electrical circuits}
\author{Motohiko Ezawa}
\affiliation{Department of Applied Physics, University of Tokyo, Hongo 7-3-1, 113-8656,
Japan}

\begin{abstract}
Non-Hermitian skin-edge states emerge only at one edge in one-dimensional
nonreciprocal chains, where all states are localized at the edge
irrespective of eigenvalues. The bulk topological number is the winding
number associated with the complex energy spectrum, which is well defined
for metals. We study non-Hermitian nonreciprocal systems in higher
dimensions, and propose to realize them with the use of electric diode
circuits. We first investigate one-dimensional interface states between two
domains carrying different topological numbers, where all states are
localized at the interface. They are a generalization of the skin-edge
states. Then we generalize them into higher dimensions. We show that there
emerge a rich variety of boundary states and interface states including
surface, line and point states in three-dimensional systems. They emerge at
boundaries of several domains carrying different topological numbers. The
resulting systems are the first-order, second-order and third-order
topological metals. Such states may well be observed by measuring the
two-point impedance in diode circuits.
\end{abstract}

\maketitle

\textit{Introduction:} The bulk-boundary correspondence has played an
important role in topological systems. The emergence of boundary states is
expected in all boundaries of a topological insulator (TI). For instance,
when we consider a sufficiently long one-dimensional (1D) topological chain,
edge states must emerge at both sides of the chain. Recently, non-Hermitian
topological systems attract growing attention\cite%
{Bender,Bender2,Kohmoto,Scho,Weimann,Liang,Pan,Nori,Zhu,Konotop,Fu,UedaPRX,Gana,Katsura,Yuce,LangWang,UedaHOTI,EzawaLCR}%
. In this context, a non-Hermitian bulk-edge correspondence was discovered
in some 1D systems with the emergence of skin states\cite%
{Xiong,Mart,UedaPRX,Kunst,Yao,Lee,Jin,Research,SkinTop,Luo}. In particular,
boundary states emerge only at one of the edges in a nonreciprocal 1D chain.
One may wonder why the usual bulk-boundary correspondence is not valid. It
is because the system is not an insulator but a non-Hermitian metal, though
the topological charge is well defined to the bulk\cite{UedaPRX}.

These non-Hermitian problems have been investigated by employing photonic
systems\cite{Mark,Scho,Pan,Weimann}, microwave resonators\cite{Poli}, wave
guides\cite{Zeu}, quantum walks\cite{Rud,Xiao}, cavity systems\cite{Hoda}
and electric circuits\cite{EzawaLCR,Luo}. In particular, it would be
convenient to tune system parameters and to induce topological phase
transitions in electric circuits\cite%
{TECNature,ComPhys,Garcia,Hel,EzawaTEC,EzawaLCR,Luo}.

On the other hand, the bulk-boundary correspondence has been generalized to
include higher-order topological insulators\cite%
{Peng,Lang,Song,Bena,Schin,EzawaKagome,PhosHOTI}. For three-dimensional (3D)
crystals, the second-order TI has 1D hinge states but has no 2D surface
states, while the third-order TI has 0D corner states but has neither
surface nor hinge states. It is an intriguing problem how these higher-order
topological states are generalized to nonreciprocal systems.

The aim of this paper is to explore the nonreciprocal bulk-boundary
correspondence in higher dimensions. First of all, we are able to define
non-Hermitian winding numbers in 2D square and 3D cubic systems, which are
valid for gapless systems. Second, we obtain higher-order topological metals
together with a rich variety of topological boundary and interface states.
These states emerge together with characteristic peaks in the local density
of states (LDOS). We also propose a method to detect such states by
measuring two-point impedance in appropriately designed electric circuits.

\begin{figure}[t]
\centerline{\includegraphics[width=0.45\textwidth]{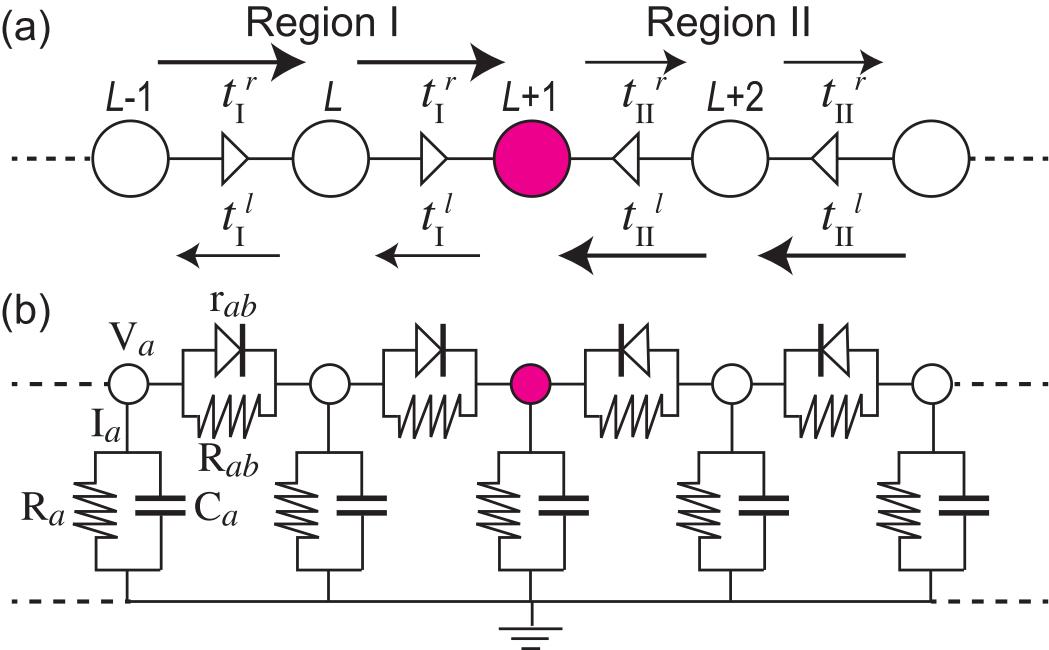}}
\caption{(a) Illustration of an interface in a 1D nonreciprocal system,
where the amplitudes are different between the leftward ($t^{r}$) and
rightward ($t^{l}$) hopping. We call the link rightward (denoted by $\rhd $)
when $|t^{r}/t^{l}|>1$ and leftward (denoted by $\lhd $) when $%
|t^{l}/t^{r}|>1$. Links are oriented in this sense. The region made of the
leftward (rightward) links has the topological number $w=+1$ ($-1$). There
emerges a topological interface state at the junction ($\rhd,\lhd$) but not
at ($\lhd,\rhd$). (b) Realization of a topological interface state by way of
a diode circuit. The topological interface state is detected by observing a
prominent impedance peak as in Fig.\protect\ref{FIG2}(a3)--(c3). }
\label{FIG1}
\end{figure}

\textit{Nonreciprocal system:} We investigate a nonreciprocal system in $n$D
hypercubic lattices [Fig.\ref{FIG1}(a)] described by 
\begin{equation}
H=\sum_{a<b}(t_{ab}c_{b}^{\dagger }c_{a}+t_{ba}c_{a}^{\dagger
}c_{b})+\sum_{a}U_{a}c_{a}^{\dagger }c_{a},  \label{HamilTB}
\end{equation}
where $t_{ab}$ is the hopping amplitude between adjacent sites and $U_{a}$
is the on-site potential. Sites $a$ and $b$ are on the same axis in a
hypercubic lattice, and the condition $a<b$ is well defined. The system is
non-Hermitian for $\left\vert t_{ab}\right\vert \neq \left\vert
t_{ba}\right\vert $. The potential $U_{a}$ can be complex. The 1D model is
known as the Hatano-Nelson model\cite{Hatano,UedaPRX}. In the homogeneous
system where $t_{ab}=t_{\mu }^{r}$, $t_{ba}=t_{\mu }^{l}$ for $a<b$ and $%
U_{a}=U$, the energy is given by 
\begin{equation}
E\left( \mathbf{k}\right) =\sum_{\mu }(t_{\mu }^{r}e^{-ik_{\mu }}+t_{\mu
}^{l}e^{ik_{\mu }})+U,
\end{equation}
where $\mu =x$ for 1D, $\mu =x,y$ for 2D, and $\mu =x,y,z$ for 3D.

The energy is a complex number, which allows us to define a non-Hermitian
winding number in $n$D as 
\begin{equation}
w_{\mu }=\int_{-\pi }^{\pi }\frac{dk_{\mu }}{2\pi i}\partial _{k_{\mu }}\ln %
\left[ E\left( \mathbf{k}\right) -\bar{E}_{\mu }\right] ,  \label{2DWind}
\end{equation}%
where we have defined 
\begin{equation}
\bar{E}_{\mu }=\int_{-\pi }^{\pi }\frac{dk_{\mu }}{2\pi }E\left( \mathbf{k}%
\right) .
\end{equation}%
We note that $E\left( \mathbf{k}\right) -\bar{E}_{\mu }$ only depends on $%
k_{\mu }$ since $U_{\mu }$ gives a constant term with respect to $k_{\mu }$.
It agrees with the previous definition\cite{Fu,UedaPRX} for 1D, where $%
w_{\mu }=1$ for $\left\vert t_{\mu }^{r}\right\vert <\left\vert t_{\mu
}^{l}\right\vert $ and $w_{\mu }=-1$ for $\left\vert t_{\mu }^{r}\right\vert
>\left\vert t_{\mu }^{l}\right\vert $. It is understood as the vorticity of
the energy\cite{Fu}.

\begin{figure}[t]
\centerline{\includegraphics[width=0.49\textwidth]{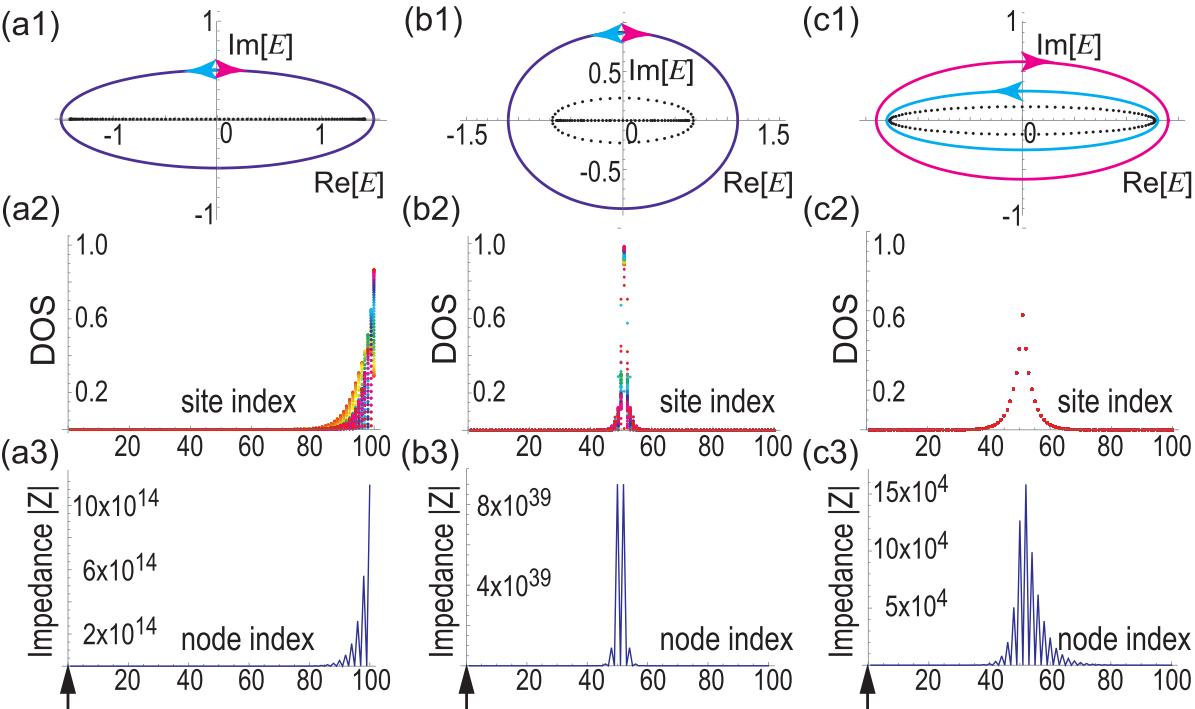}}
\caption{(a1)--(c1) Energy spectrum $E(k)$ in the complex plane, where $%
0\leq k\leq 2\protect\pi $. The horizontal axis is Re[$E$] and the vertical
axis is Im[$E$]. As $k$ increases from $k=0$, $E(k)$ draws a magenta (cyan)
oval to form the bulk spectrum for the region I (II) and black dots to form
the interface spectrum. A magenta (cyan) oval encircles the origin once
clockwise (anticlockwise), yielding the topological number $w=-1(+1)$ to the
region. Magenta and cyan ovals are degenerate in (a1) and (b1). (a2)--(c2)
Plot of the LDOS for all eigenvalues, where different eigenvalues are
indicated by different colors. All eigenvalues are degenerate in (c2).
(a3)--(c3) Two-point impedance ($|Z|$ in unit of $\Omega $) where one node
is fixed at the edge node indicated by an upward arrow. (a1)--(a3) for the
edge state, where we set $t^{r}=1$, $t^{l}=1/2$. (b1)--(b3) for the
interface state at the junction ($\rhd ,\lhd $) as in Fig.\protect\ref{FIG1}%
, where we set $t_{\text{l}}^{r}=t_{\text{lI}}^{l}=1$ and $t_{\text{l}%
}^{l}=t_{\text{lI}}^{r}=0.1$. (c1)--(c3) for a similar interface state,
where $t_{\text{l}}^{r}=1$, $t_{\text{l}}^{l}=0.4$, $t_{\text{lI}}^{r}=0.5$
and $t_{\text{lI}}^{l}=0.8$. }
\label{FIG2}
\end{figure}

The $n$D system is characterized by a set of $n$ winding numbers. There are
two topological phases $w=\pm 1$ for 1D, there are four topological phases $%
(w_{x},w_{y})=(\pm 1,\pm 1)$ for 2D, and there are eight topological phases $%
(w_{x},w_{y},w_{z})=(\pm 1,\pm 1,\pm 1)$ for 3D. The system is always
metallic since it is a one-band system. It is notable that the topological
number is defined for metals, which is contrasted to the usual topological
systems, where the topological number is defined for gapped systems such as
topological insulators and superconductors. It is intriguing that we can
define topological numbers for one-band systems in any dimensions.

The usual bulk-edge correspondence for topological insulators implies the
emergence of the topological edges at both edges. Nevertheless, the
bulk-edge correspondence for the present model is drastically different,
which is allowed since the system is metallic: The topological edge states
emerge only at the left (right) edge when $\ w=1$ ($w=-1$) for the 1D system%
\cite{UedaPRX}. A generalization to higher dimensional systems is a highly
nontrivial problem, which we will explore in this paper.

\textit{Diode and nonreciprocal system:} We propose a method to examine the
results of the above nonreciprocal systems by realizing them with the use of
electric circuits. Typical nonreciprocal devices are diodes. We thus
consider a diode circuit as in Fig.\ref{FIG1}(b). The current $I$\ is not
proportional to the voltage $V$\ for diodes, where the resistance has a
dependence on $V$. However, we may in general approximate it by a linear
resistance $r_{ab}$\ for $a<b$. On the other hand, the resistance is perfect
nonreciprocal, $r_{ba}=\infty $\ for $a<b$. In this approximation, we obtain
a linear nonreciprocal system with constant resistors.

The Kirchhoff's current law for the circuit in Fig.\ref{FIG1}(b) is
expressed as 
\begin{equation}
I_{a}=C_{a}\frac{d}{dt}V_{a}+\frac{V_{a}}{R_{a}}+\sum_{b}\frac{V_{a}-V_{b}}{%
\bar{R}_{ab}},  \label{Kirch}
\end{equation}%
where the sum over $b$\ is taken for the two nearest neighbor nodes of $a$; $%
I_{a}$\ is the current between node $a$\ and the ground, $V_{a}$\ is the
voltage at node $a$, $R_{a}$\ and $C_{a}$\ are the resistance and the
capacitance connected in parallel to the ground; finally, $\bar{R}_{ab}$\ is
the resistance between nodes $a$\ and $b$, 
\begin{equation}
\bar{R}_{ab}=\left\{ 
\begin{array}{cc}
r_{ab}R_{ab}/(r_{ab}+R_{ab}) & \text{for\quad }a<b \\ 
R_{ab} & \text{for\quad }a>b%
\end{array}%
\right. .  \label{KirchA}
\end{equation}%
Here, we have added resistance $R_{ab}$\ between two nodes $a$\ and $b$\ in
parallel to the diode. The resistor $R_{ab}$\ modifies the perfect
nonreciprocity to imperfect nonreciprocity, while the resistor $R_{a}$\
modifies the potential term, as we shall soon see below.

\begin{figure*}[t]
\centerline{\includegraphics[width=0.96\textwidth]{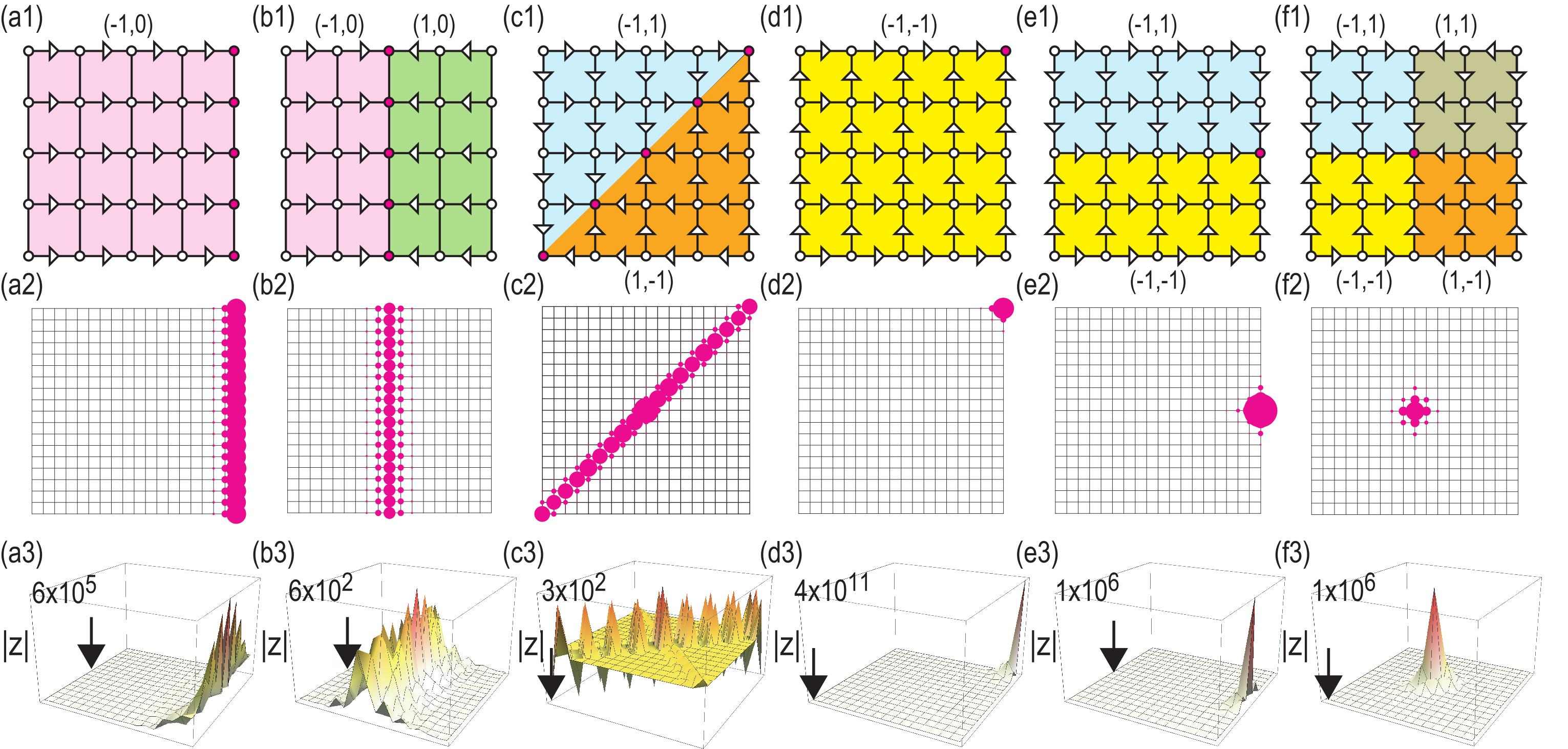}}
\caption{(a1)--(f1) Illustration of various topological boundary and
interface states (marked in magenta) in 2D nonreciprocal systems. These
states emerge at a boundary or an interface where the opposite
nonreciprocities meet in such an order as indicated in the figures. They are
(a) edge, (b) interface, (c) diagonal, (d) corner (e) edge-central and (f)
bulk-central states. We obtain first-order topological metals (a,b,c), and
second-order topological metals (d,e,f). Links are oriented in the same way
as in the 1D nonreciprocal system (see Fig.\protect\ref{FIG1}). Here in 2D,
there are four orientations, rightward ($\rhd $), leftward ($\lhd $), upward
($\triangle$) and downward ($\triangledown$). A 2D region is indexed by a
pair of topological charges ($w_x,w_y$), where $w_x=\pm1$ and $w_y=\pm1$.
(a2)--(f2) Sum of the LDOS of the eigen functions for all eigenvalues, which
have large peaks at topological boundaries: The magnitude of a peak is
represented by the size of a magenta disk. It is possible to interpret the
lattice structures (a1)--(f1) as diode circuits as in Fig.\protect\ref{FIG1}%
. (a3)--(f3) Two-point impedance ($|Z|$ in unit of $\Omega$), where one node
is fixed as indicated by an arrow. A link is either oriented (nonreciprocal)
or not (reciprocal). If not we have set $t_{ab}=1$. If oriented, we have set 
$t_{ab}=1$ for the positive orientation and $t_{ab}=1/4$ for the negative
orientation.}
\label{FIG3}
\end{figure*}

The admittance matrix $J_{ab}\left( \omega \right) $\ is defined by\cite%
{ComPhys,TECNature,Hel,Lu,EzawaTEC,EzawaLCR} 
\begin{equation}
I_{a}\left( \omega \right) =\sum_{b}J_{ab}\left( \omega \right) V_{b}\left(
\omega \right) ,
\end{equation}
with 
\begin{equation}
J_{ab}\left( \omega \right) \equiv \left( -\sum_{b}\frac{1}{\bar{R}_{ab}} +%
\frac{1}{R_{a}}+i\omega C_{a}\right) \delta _{ab} +\frac{1}{\bar{R}_{ab}}.
\end{equation}
There is a one-to-one correspondence between the admittance matrix $J_{ab}$
and the tight-binding Hamiltonian (\ref{HamilTB}) with 
\begin{equation}
t_{ab}=\frac{1}{\bar{R}_{ab}},\quad V_{a}=-\sum_{b}\frac{1}{\bar{R}_{ab}}+%
\frac{1}{R_{a}}+i\omega C_{a}.  \label{Circuit}
\end{equation}
We can set the real part of $V_{a}$ to be a constant by tuning $R_{a}$. The
perfect nonreciprocal system with the forward-backward connection is
realized by connecting perfect diodes so that the current flows right (left)
going in the region I (II), as shown in Fig.\ref{FIG1}.

The local density of states of the eigenfunction of $J_{ab}$ is well
observed by the two-point impedance, which is given by\cite%
{ComPhys,TECNature} 
\begin{equation}
Z_{ab}=\frac{V_{a}-V_{b}}{I_{ab}}=G_{aa}+G_{bb}-G_{ab}-G_{ba},  \label{EqZ}
\end{equation}
where $G$ is the Green function defined by the inverse of the admittance
matrix $G_{ab}\equiv J_{ab}^{-1}$.

\textit{1D nonreciprocal system:} We first study the Hatano-Nelson model in
order to study interface states. We consider a closed chain with the length $%
2L$, where the hoppings are $t_{\text{I}}^{r}$ and $t_{\text{I}}^{l}$ for $%
1\leq n\leq L$ and $t_{\text{II}}^{r}$ and $t_{\text{II}}^{l}$ for $L+1\leq
n\leq 2L$. We call the sites $1\leq n\leq L$ the region I and the sites $%
L+1\leq n\leq 2L$ the region II. Being a closed chain, it has two interfaces
at the sites $n=L+1$ and $n=1$.

\begin{figure*}[t]
\centerline{\includegraphics[width=0.9\textwidth]{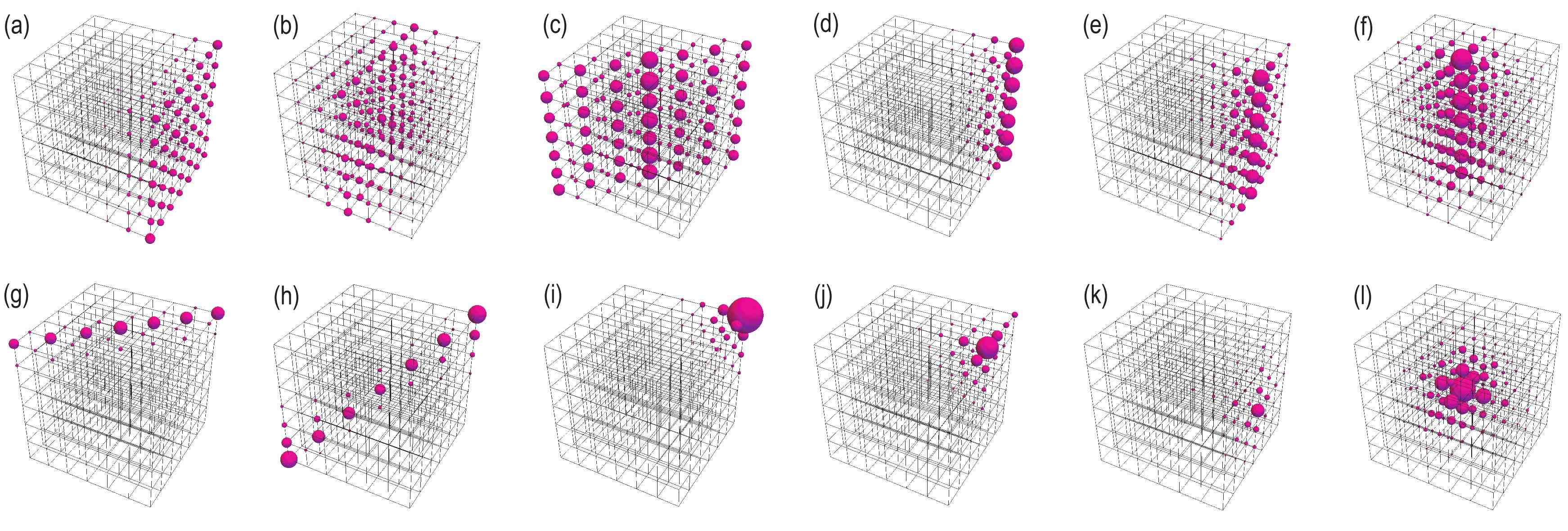}}
\caption{Illustration of various topological boundary and interface states
and LDOS in 3D nonreciprocal systems. The magnitude of the LDOS peak is
represented by the size of the magenta ball. These states emerge at a
boundary or an interface where the opposite nonreciprocities meet in a
certain order. LDOS representing (a) surface, (b) interface, (c) diagonal
interface, (d) hinge, (e) surface-central line, (f) body-central line, (g)
surface-diagonal line, (h) body-diagonal line and (i) corner, (j)
hinge-central point, (k) surface-central point and (l) body-central point
states. They are first-order topological metals (a,b,c), second-order
topological metals (d,e,f,g,h), and third-order topological metals
(i,j,k,l). A link is either oriented (nonreciprocal) or not (reciprocal). If
not we have set $t_{ab}=1$. If oriented, we have set $t_{ab}=1$ for the
positive orientation and $t_{ab}=1/4$ for the negative orientation. }
\label{FIG4}
\end{figure*}

In order to study non-Hermitian interface states, we diagonalized the
Hamiltonian (\ref{HamilTB}) in 1D. The eigen function is indexed by the site 
$n$ and labeled as $\left( \psi _{1},\psi _{2},\cdots \right) $. We may
solve the eigenvalue problem exactly when the relation $t_{\text{I}}^{l}t_{%
\text{I}}^{r}=t_{\text{II}}^{l}t_{\text{II}}^{r}$ and $U_{a}=0$ hold. For $%
\left\vert t_{\text{II}}^{r}/t_{\text{I}}^{r}\right\vert <1$, the interface
states emerge only at $n=L+1$ and are given by\cite{SupMat}
\begin{eqnarray}
\psi _{n} &=&\alpha ^{L+1-n}\text{ for }n\leq L, \\
\psi _{n} &=&\beta ^{n-L-1}\text{ for }n>L,
\end{eqnarray}
with $\alpha =\sqrt{t_{\text{II}}^{r}/t_{\text{I}}^{r}}e^{i\pi /L}$ and $%
\beta =\sqrt{t_{\text{II}}^{r}/t_{\text{I}}^{r}}e^{-i\pi /L}$. The
eigenvalues are $E=\sqrt{t_{\text{I}}^{r}t_{\text{II}}^{r}}e^{i\pi /L}+\sqrt{%
t_{\text{I}}^{l}t_{\text{II}}^{l}}e^{-i\pi /L}$. All eigen functions are
exponentially localized at the site $n=L+1$. All penetration depths are
identical and given
by $\xi =\frac{1}{2}\left\vert \ln \left( t_{\text{I}}^{r}/t_{\text{II}}^{r}\right) \right\vert $ between the regions I and II.
%See the derivation in Supplemental Material.

There is another exact solution. When $t_{\text{I}}^{r}\neq t_{\text{II}%
}^{l} $, $t_{\text{I}}^{r}\neq 0$, $t_{\text{II}}^{l}\neq 0$ and $t_{\text{I}%
}^{l}=t_{\text{II}}^{r}=0$ and $U_{n}=0$, all eigenvalues are zero and only
the two eigen functions are nontrivial, 
\begin{equation}
\psi _{n}^{\left( 1\right) }\propto -t_{\text{II}}^{l}\delta _{n,L}+t_{\text{%
I}}^{r}\delta _{n,L+2}, \qquad \psi _{n}^{\left( 2\right) }=\delta _{n,L+1}.
\end{equation}
The localization is strict and there is no penetration depth.

We may carry out numerical diagonalizations for any set of parameters. The
bulk topological number is $w=-1$ in the region I and $w=1$ in the region
II, which implies that the emergence of skin states at the right edge of the
region I and the left edge of the region II. It is consistent with the fact
that the interface states emerge at the site $n=L+1$ but there are no
interface states at one site $n=1$. They are topological nonreciprocal
interface states.

We show the energy spectrum and the LDOS in Fig.\ref{FIG2}. We have
calculated the two-point impedance based on the formula (\ref{EqZ}) and show
the results also in Fig.\ref{FIG2}. The characteristic behaviors of the LDOS
are well signaled by measuring the two-point impedance, where one node is
fixed at the site $n=1$ as in Fig.\ref{FIG2}(a3)--(c3).

\textit{2D square nonreciprocal system:} We proceed to study the
nonreciprocal Hamiltonian (\ref{HamilTB}) in 2D. The simplest one is just to
stack finite 1D chains, where the hopping along the $y$-axis is reciprocal.
The bulk topological number is ($\pm 1,0$). Skin-edge states emerge along
the $y$-axis as in Fig.\ref{FIG3}(a1). The next simplest one is to stack
finite 1D chains with an interface. We obtain interface states along the $y$%
-axis at the junction of the two phases with ($-1,0$) and ($1,0$) in this
order as in Fig.\ref{FIG3}(b1). It is also possible to have a diagonal
interface state formed at the junction of the two phases with ($-1,-1$) and (%
$1,1$) in this order as in Fig.\ref{FIG3}(c1). These 1D objects are
first-order topological boundary or interface states in 2D.

We may also have second-order topological boundary or interface states in
2D, which are 0D objects. Nonreciprocal corner states emerge at the right-up
corner for the phase with ($-1,-1$) as in Fig.\ref{FIG3}(d1). Similarly, we
have the right-down corner for the phase with ($-1,1$), the left-up corner
for the phase with ($1,-1$) and the left-down corner for the phase with ($%
1,1 $). Here we note a previous report\cite{SkinTop} showing the emergence
of a corner state at the right-up corner without any bulk topological
numbers. Furthermore, we obtain edge-central states at the corners of two
phases with $\left( -1,1\right) $ and $\left( -1,-1\right) $, as showing Fig.%
\ref{FIG3}(e). Finally, we have bulk-central states at the corners of four
phases with $\left( -1,1\right) $, $\left( 1,1\right) $, $\left(
-1,-1\right) $ and $\left( 1,-1\right) $, as shown in Fig.\ref{FIG3}(f).

These boundary or interface states are manifested by evaluating the LDOS,
whose results are shown in Fig.\ref{FIG3}(a2)--(f2). Experimental
verifications would be carried out by designing diode circuits implied by
Fig.\ref{FIG3}(a1)--(f1). We have calculated the two-point impedance based
on the formula (\ref{EqZ}) by fixing one node as indicated by an arrow: See
Fig.\ref{FIG3}(a3)--(f3).  The behavior of the two-point impedance is not so
simple as that of the LDOS, because it depends on the fixed point rather
sensitively.

\textit{3D cubic nonreciprocal system:} We also study the nonreciprocal
Hamiltonian (\ref{HamilTB}) in 3D. The simplest ones are just to stack 2D
squares, where the hopping along the $z$-axis is reciprocal. For instance,
we obtain 2D surface boundary or interface states Fig.\ref{FIG4}(a), (b) and
(c) from Fig.\ref{FIG3}(a1), (b1) and (c1), respectively. They constitute
the first-order topological metals. Similarly we obtain 1D line states Fig.%
\ref{FIG4}(d), (e) and (f) at a boundary or interface from Fig.\ref{FIG4}%
(d1), (e1) and (f1), respectively. We may have another types of 1D lines as
in Fig.\ref{FIG4}(g), (h) when appropriate nonreciprocity is introduced into
the $z$-axis. They constitute the second-order topological metals. Finally,
we may make a full use of nonreciprocity to generate 0D point states as in
Fig.\ref{FIG4}(i), (j), (k) and (l). The bulk of the corner state Fig.\ref%
{FIG4}(i) has the topological charge $\left( -1,-1,-1\right) $. The
bulk-central point in Fig.\ref{FIG4}(l) emerges at the common corners of
eight topological phases $\left( w_{x},w_{y},w_{z}\right) $ where $w_{x}=\pm
1$, $w_{y}=\pm 1$ and $w_{z}=\pm 1$. They constitute the third-order
topological metals.

The author is very much grateful to N. Nagaosa for helpful discussions on
the subject. This work is supported by the Grants-in-Aid for Scientific
Research from MEXT KAKENHI (Grants No. JP17K05490, No. JP15H05854 and No.
JP18H03676). This work is also supported by CREST, JST (JPMJCR16F1 and
JPMJCR1874).

\newpage

\def\theequation{S\arabic{equation}}
\def\thefigure{S\arabic{figure}}
\def\thesubsection{S\arabic{subsection}}
\setcounter{equation}{0}

%\begin{widetext}
%\onecolumn
\begin{center}\textbf{\large Supplemental Material}\\[10pt] 
\end{center}
%\section*{Non-Hermitian boundary and interface states in nonreciprocal higher-order topological metals and electrical circuits}
%\end{widetext}

We derive eqs.(11) and (12) in the main text. We write the eigen function as 
$\left( \psi _{1},\psi _{2},\cdots ,\psi _{n},\cdots \right) $ with the site
index $n$. The eigen equation is given by%
\begin{equation}
\left( 
\begin{array}{ccccccccc}
-\varepsilon  & t_{\text{I}}^{l} & 0 & \cdots  & \cdots  & \cdots  & \cdots 
& 0 & t_{\text{II}}^{r} \\ 
t_{\text{I}}^{r} & -\varepsilon  & t_{\text{I}}^{l} & 0 & \cdots  & \cdots 
& \cdots  & \cdots  & 0 \\ 
\ddots  & \ddots  & \ddots  & \ddots  & \ddots  & \ddots  & \ddots  & \ddots 
& \ddots  \\ 
0 & \cdots  & t_{\text{I}}^{r} & -\varepsilon  & t_{\text{I}}^{l} & 0 & 0 & 
\cdots  & 0 \\ 
0 & \cdots  & 0 & t_{\text{I}}^{r} & -\varepsilon  & t_{\text{II}}^{l} & 0 & 
\cdots  & 0 \\ 
0 & \cdots  & 0 & 0 & t_{\text{II}}^{r} & -\varepsilon  & t_{\text{II}}^{l}
& \cdots  & 0 \\ 
\ddots  & \ddots  & \ddots  & \ddots  & \ddots  & \ddots  & \ddots  & \ddots 
& \ddots  \\ 
0 & \cdots  & \cdots  & \cdots  & \cdots  & 0 & t_{\text{II}}^{r} & 
-\varepsilon  & t_{\text{II}}^{l} \\ 
t_{\text{II}}^{l} & 0 & \cdots  & \cdots  & \cdots  & \cdots  & 0 & t_{\text{%
II}}^{r} & -\varepsilon 
\end{array}%
\right) \left( 
\begin{array}{c}
\psi _{1} \\ 
\psi _{2} \\ 
\cdots  \\ 
\psi _{L} \\ 
\psi _{L+1} \\ 
\psi _{L+2} \\ 
\cdots  \\ 
\psi _{2L-1} \\ 
\psi _{2L}%
\end{array}%
\right) =0.
\end{equation}%
We assume exponentially decaying solutions,%
\begin{equation}
\psi _{L+1-n}=\alpha ^{n},\qquad \psi _{L+1+n}=\beta ^{n}.  \label{0}
\end{equation}%
The eigen equation is simplified as%
\begin{eqnarray}
t_{\text{I}}^{r}\alpha ^{2}-\varepsilon \alpha +t_{\text{I}}^{l} &=&0,
\label{1} \\
t_{\text{I}}^{r}\alpha -\varepsilon +t_{2}^{l}\beta  &=&0,  \label{2} \\
t_{\text{II}}^{r}-\varepsilon \beta +t_{\text{II}}^{l}\beta ^{2} &=&0.
\label{3}
\end{eqnarray}%
We solve eq.(\ref{2}) as%
\begin{equation}
\varepsilon =t_{\text{I}}^{r}\alpha +t_{\text{II}}^{l}\beta .  \label{4}
\end{equation}%
By inserting eq.(\ref{4}) to eqs. (\ref{1}) and (\ref{3}), we find%
\begin{eqnarray}
-t_{\text{II}}^{l}\alpha \beta +t_{\text{I}}^{l} &=&0, \\
t_{\text{II}}^{r}-t_{\text{I}}^{r}\alpha \beta  &=&0,
\end{eqnarray}%
which gives the relation%
\begin{equation}
\alpha \beta =\frac{t_{\text{I}}^{l}}{t_{\text{II}}^{l}}=\frac{t_{\text{II}%
}^{r}}{t_{\text{I}}^{r}},
\end{equation}%
or 
\begin{equation}
t_{\text{I}}^{l}t_{\text{I}}^{r}=t_{\text{II}}^{l}t_{\text{II}}^{r},
\end{equation}%
which is the condition given in the main text. Since the chain is closed, we
also have the equations 
\begin{eqnarray}
t_{\text{II}}^{r}\psi _{2L}-\varepsilon \psi _{1}+t_{\text{I}}^{l}\psi _{2}
&=&0, \\
t_{\text{II}}^{r}\psi _{2L-1}-\varepsilon \psi _{2L}+t_{\text{II}}^{l}\psi
_{1} &=&0.
\end{eqnarray}%
By inserting eq.(\ref{0}) into these, we obtain%
\begin{eqnarray}
t_{\text{II}}^{r}\beta ^{L-1}-\varepsilon \alpha ^{L}+t_{\text{I}}^{l}\alpha
^{L-1} &=&0, \\
t_{\text{II}}^{r}\beta ^{L-2}-\varepsilon \beta ^{L-1}+t_{\text{II}%
}^{l}\alpha ^{L} &=&0.
\end{eqnarray}%
By using eq.(\ref{4}), we have%
\begin{equation}
\frac{1}{\beta }=\alpha \frac{t_{\text{I}}^{r}}{t_{\text{II}}^{r}}.
\end{equation}%
Finally, we obtain%
\begin{equation}
\alpha =\sqrt{t_{\text{II}}^{r}/t_{\text{I}}^{r}}e^{i\pi /L},\qquad \beta =%
\sqrt{t_{\text{II}}^{r}/t_{\text{I}}^{r}}e^{-i\pi /L},
\end{equation}%
with the eigenvalues%
\begin{equation}
\varepsilon =\sqrt{t_{1}^{\text{r}}t_{2}^{\text{r}}}e^{i\pi /L}+\sqrt{t_{1}^{%
\text{l}}t_{2}^{\text{l}}}e^{-i\pi /L}.
\end{equation}%

\end{document}